\documentclass[aps,prb,twocolumn,floatfix]{revtex4-2}
\usepackage{placeins}
\usepackage{graphicx}
\usepackage{multirow}
\usepackage{color}
\usepackage{amsmath}
\usepackage{amsfonts}
\usepackage{amssymb}
\usepackage{graphicx}
\graphicspath{{Figures/}}
\usepackage{epstopdf}
\usepackage{empheq}
\usepackage{float}
\usepackage{cases} 
\usepackage{mathtools}
\usepackage{dcolumn} 
\usepackage{bbold}
\usepackage{xfrac}
\usepackage{bm}
\usepackage{siunitx}

\usepackage[dvipsnames]{xcolor}
\usepackage[colorlinks]{hyperref}
\hypersetup{
colorlinks=True,
linkbordercolor=White,
linkcolor=BrickRed,
citecolor=Blue,	 
}

\renewcommand{\vec}[1]{\mathbf{#1}}

\begin{document}

\title{Hole spin qubits in unstrained Germanium layers}

\author{Lorenzo Mauro}
\affiliation{Univ. Grenoble Alpes, CEA, IRIG-MEM-L\_Sim, Grenoble, France.}%
\author{Mauricio J. Rodr\'iguez}
\affiliation{Univ. Grenoble Alpes, CEA, IRIG-MEM-L\_Sim, Grenoble, France.}%
\author{Esteban A. Rodr\'iguez-Mena}
\affiliation{Univ. Grenoble Alpes, CEA, IRIG-MEM-L\_Sim, Grenoble, France.}%
\author{Yann-Michel Niquet}
\email{yniquet@cea.fr}
\affiliation{Univ. Grenoble Alpes, CEA, IRIG-MEM-L\_Sim, Grenoble, France.}%

\date{\today}

\begin{abstract}
Strained germanium heterostructures are one of the most promising material for hole spin qubits but suffer from the strong anisotropy of the gyromagnetic factors that hinders the optimization of the magnetic field orientation. The figures of merit (Rabi frequencies, lifetimes...) can indeed vary by an order of magnitude within a few degrees around the heterostructure plane. We propose to address this issue by confining the holes at the interface of an unstrained, bulk Ge substrate or thick buffer. We model such structures and show that the gyromagnetic anisotropy is indeed considerably reduced. In addition, the Rabi frequencies and quality factors can be significantly improved with respect to strained heterostructures. This extends the operational range of the qubits and shall ease the scale-up to many-qubit systems.
\end{abstract}

\maketitle

\section{Introduction}

Hole spin qubits in semiconductor quantum dots have made remarkable progress as a compelling platform for quantum computing and simulation \cite{Loss98,Maurand16,Froning21,Camenzind22,wang2022ultrafast,Fang2023Review}. One of their main assets is the efficient electrical manipulation enabled by the strong spin-orbit coupling (SOC) in the valence bands of semiconductor materials \cite{Winkler03,Rashba03,Kato03,Golovach06,Crippa18}. In particular, planar germanium heterostructures now stand out as the state-of-the-art material for hole spin qubits \cite{Sammak19,Scappucci20}. The quality of epitaxial interfaces indeed reduces the disorder around the qubits \cite{Martinez2022,Varley23,Massai23,Martinez24}, and the small effective mass of holes in Ge allows for larger quantum dots, which eases fabrication and integration. High-fidelity single and two-qubit gates have thus been reported in germanium heterostructures \cite{Watzinger18,Hendrickx20b,Hendrickx20,Hendrickx21,Lawrie23,Hendrickx2024}, in up to ten qubits \cite{Valentin25}. Singlet-triplet spin qubits \cite{jirovec_Freqs_ST-Ge,Jirovec23,Zhang25,Tsoukalas2025}, manipulation by spin shuttling \cite{Wang2024}, and quantum simulation \cite{Wang2023} have also been demonstrated on this platform.

In these heterostructures, the heavy-hole (HH) and light-hole (LH) subbands are strongly split by the biaxial strains resulting from the growth on a mismatched GeSi buffer \cite{Sammak19}. As a consequence, the low-lying hole states have strong HH character, and thus show a highly anisotropic gyromagnetic response, with in-plane $g$-factors $g_\parallel\lesssim 0.5$ and out-of-plane $g$-factors $g_\perp\gtrsim 10$ \cite{Hendrickx2024}. Therefore, all relevant spin properties (Larmor and Rabi frequencies, lifetimes, ...) vary rapidly (over $\approx 1^\circ$) when the magnetic field crosses the heterostructure plane where these devices are usually operated \cite{Mauro24}. This can hinder the optimal alignment of the magnetic field, especially in many-qubit systems with significant variability. Moreover, the small HH/LH mixing limits the maximum Rabi frequencies achieved in these devices.

It would, therefore, be desirable to increase the HH/LH mixing and reduce the $g$-factor anisotropy. As discussed in Ref.~\cite{Mauro25}, this may be achieved with strain engineering, but a scalable design is still lacking. An alternative solution is to host the qubits in a bulk Ge substrate insulated from the gate stack by a thin, strained GeSi barrier \cite{Stehouwer23,Patent}. The quantum dots are then accumulated at the Ge/GeSi interface by the electric field from the gates. As the Ge substrate is unstrained, the HH/LH band gap is expected significantly smaller and the HH/LH mixing much stronger \cite{Bosco21b}. The growth of such a structure, with the formation of a high mobility hole gas at the interface, has actually been demonstrated very recently \cite{Scappucci25}.

In this work, we explore the prospects for unstrained bulk Ge qubits with detailed numerical simulations. We analyze the dependence of the $g$-factors on the electrical confinement, and show that the $g$-factor anisotropy can indeed be significantly reduced even for moderate HH/LH mixings. The stronger mixing increases the Rabi frequencies $f_\mathrm{R}$ but decreases the dephasing time $T_2^*$; nonetheless, the quality factor $Q_2^*=2f_\mathrm{R}T_2^*$ is larger than in strained heterostructures. Most importantly, the dependence of these quantities on the magnetic field orientation is much broadened, allowing for an easier optimization of the operating point in many-qubit systems. We discuss the implications for the development of hole spin qubit technologies.

\section{Results}

\subsection{Device and methodology}

\begin{figure}[t]
\centering
\includegraphics[width=0.8\linewidth]{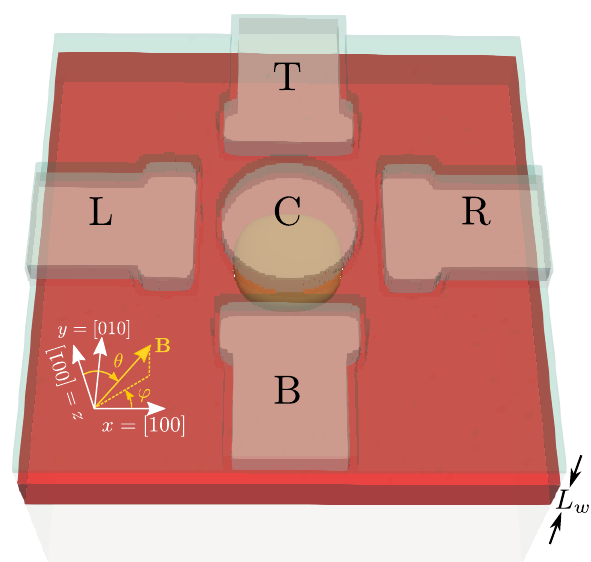}
\caption{The test device is made of a Ge well (red) with thickness $L_w$ ranging from 10\,nm to $L_w\to\infty$ (bulk). It is capped with a 20-nm-thick Ge$_{0.8}$Si$_{0.2}$ barrier (blue). The dot is shaped by five Al gates (gray) embedded in 5 nm of Al$_2$O$_3$. The diameter of the central gate is $d=100$\,nm. The yellow shape illustrates the location and shape of the quantum dot. The orientation of the magnetic field $\vec{B}$ is characterized by the angles $\theta$ and $\varphi$ in the crystallographic axes set $x=[100]$, $y=[010]$ and $z=[001]$.}
\label{fig:device}
\end{figure}

In order to compare strained and unstrained Ge qubits, we consider the same prototypical device as in Refs. \cite{martinez2022hole,Abadillo2023} (see Fig.~\ref{fig:device}). The heterostructure comprises a Ge well with thickness $L_w$ laid on Ge$_{0.8}$Si$_{0.2}$ and capped with a 20-nm-thick Ge$_{0.8}$Si$_{0.2}$ barrier. The difference of potential between the central C gate and the side L, R, T, and B gates on top of the heterostructure shapes a quantum dot in the Ge well. We address two hypotheses: {\it i}) the whole heterostructure is grown coherently on a thick Ge$_{0.8}$Si$_{0.2}$ buffer with a small, residual in-plane strain $\varepsilon_\mathrm{buf}=0.26\%$. The Ge well then undergoes compressive biaxial strains $\varepsilon_{xx}=\varepsilon_{yy}=\varepsilon_\parallel=-0.61\%$ and $\varepsilon_{zz}=\varepsilon_\perp=0.45\%$. This case is representative of the experimental, strained Ge heterostructures \cite{Sammak19}; {\it ii}) the Ge well is unstrained but the Ge$_{0.8}$Si$_{0.2}$ layers undergo tensile strains $\varepsilon_{xx}=\varepsilon_{yy}=0.87\%$ and $\varepsilon_{zz}=-0.66\%$. The bulk germanium device is the limit $L_w\to\infty$ \footnote{The bulk germanium device is practically modeled as $L_w=170$\,nm.}. Although finite $L_w$'s may hardly be grown, they provide valuable insights into the physics of the device. In both configurations, the barrier at the Ge/GeSi interface is $\Delta E_\mathrm{HH}\approx 140$\,meV for HH states and $\Delta E_\mathrm{LH}\approx 80$\,meV for LH states.

The spin of a single hole trapped in this quantum dot is manipulated with radio-frequency signals applied either to the central or side gates. We compute the potential created by the gates with a finite volumes Poisson solver, then the wave function of the hole with a finite-differences discretization of the Luttinger-Kohn Hamiltonian \cite{martinez2022hole,Abadillo2023,Luttinger56,KP09}. We finally calculate the Larmor and Rabi frequencies of the spin with the $g$-matrix formalism \cite{Venitucci18}. We do not account here for the inhomogeneous strains imprinted by the differential thermal contraction of the materials when the device is cooled down \cite{Abadillo2023}. The latter are discussed in the Supplementary Information.

\subsection{Dimensions and $g$-factors of the dots}

The effective Hamiltonian of the hole spin can be written $H=\frac{1}{2}\mu_B\boldsymbol{\sigma}\cdot\hat{g}\vec{B}$, where $\mu_B$ is Bohr's magneton, $\vec{B}$ is the magnetic field, $\boldsymbol{\sigma}$ is the vector of Pauli matrices and $\hat{g}$ is the gyromagnetic matrix \cite{Venitucci18}. For quantum dots with quasi-circular symmetry, this matrix is diagonal, with principal $g$-factors $g_{xx}=-g_{yy}=g_\parallel$, and $g_{zz}=g_\perp$ \cite{martinez2022hole,Abadillo2023}.

\begin{figure}[t]
\centering
\includegraphics[width=\linewidth]{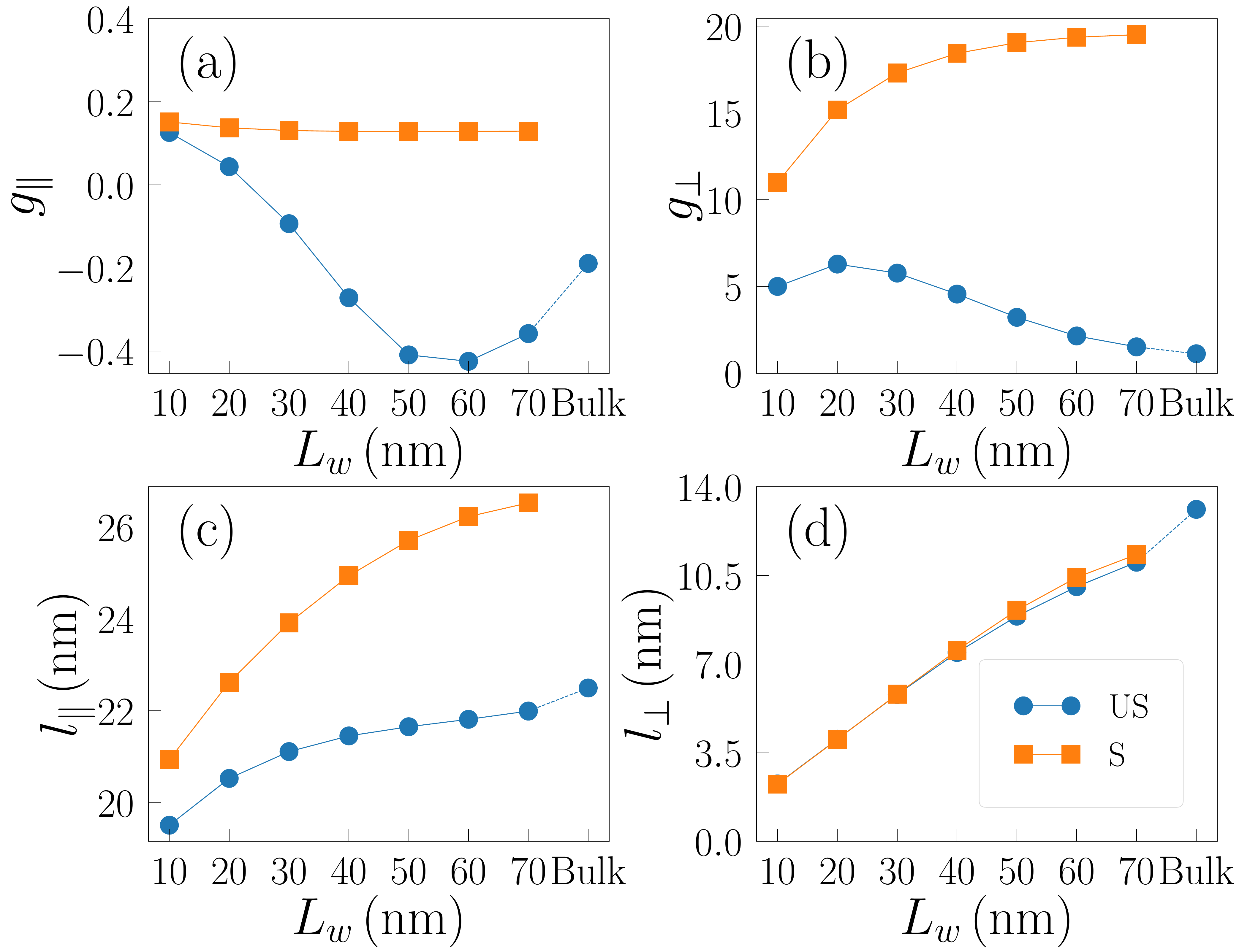}
\caption{(a)-(b) $g$-factors $g_\parallel$ and $g_\perp$ as a function of the thickness $L_w$ of strained (S) and unstrained (US) Ge wells. (c)-(d) In-plane extension $l_\parallel$ and out-of-plane extension $l_\perp$ of the dot as a function of $L_w$. All calculations are performed at $V_\mathrm{C}=-10$\,mV with the side gates grounded.}  
\label{fig:gfactLw}
\end{figure}

The $g$-factors $g_\parallel$ and $g_\perp$ are plotted as a function of the well thickness $L_w$ in Fig.~\ref{fig:gfactLw} for both strained and unstrained Ge wells. The potential applied to the central gate is $V_\mathrm{C}=-10$\,mV and the side gates are grounded. The vertical extension $\ell_\perp=\sqrt{\langle z^2\rangle-\langle z\rangle^2}$ and the lateral extension $\ell_\parallel=\sqrt{\langle x^2\rangle}=\sqrt{\langle y^2\rangle}$ of the dot are also shown in this figure.

As expected, the vertical extension of the dot increases with $L_w$, but is ultimately limited by the vertical electric field from the gates that tends to squeeze the hole at the top Ge/Ge$_{0.8}$Si$_{0.2}$ interface. Indeed, $\ell_\perp\approx 0.18L_w$ is consistent with a square well model for small $L_w$, but departs from this trend in thick Ge films. There is little difference between strained and unstrained wells as the biaxial strain has almost no effect on the vertical confinement mass. The dot also extends laterally with increasing $L_w$, primarily because the in-plane (transport) mass of the hole decreases (from $m_\parallel\approx 0.08\,m_0$ for $L_w=10$\,nm to $m_\parallel\approx 0.06\,m_0$ for $L_w=70$\,nm, with $m_0$ the free electron mass) \footnote{To lowest order, the transport mass is $m_\parallel\approx m_0/(\gamma_1+\gamma_2-\gamma_h)$, where $\gamma_1=13.18$, $\gamma_2=4.24$ and $\gamma_h$ is defined after Eq.~\eqref{eq:gHH} \cite{Michal21}.}. This is also why the dots are more localized in the unstrained Ge wells that show heavier transport masses ($m_\parallel\approx 0.1\,m_0$ for $L_w=10$\,nm and $m_\parallel\approx 0.08\,m_0$ for $L_w=70$\,nm). Although undesirable, the enhanced localization in unstrained Ge wells remains limited.

\begin{figure}[t]
\centering
\includegraphics[width=0.75\linewidth]{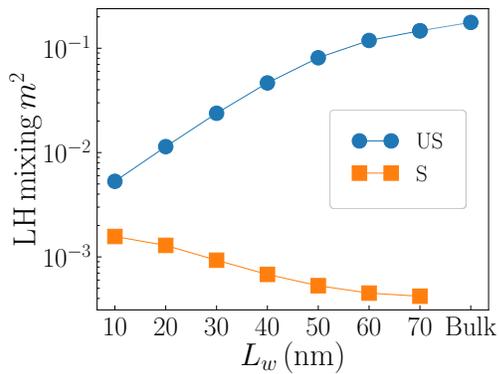}
\caption{Weight $m^2$ of the LH components in the ground-state hole wave function, as a function of the thickness $L_w$ of strained (S) and unstrained (US) Ge wells.} 
\label{fig:LHmix}
\end{figure}

The gyromagnetic factors follow very different trends in strained and unstrained Ge wells. In particular, the $g$-factor anisotropy $g_\perp/g_\parallel$ is much reduced in bulk Ge, as a result of the enhancement of the HH/LH mixing. Indeed, the $g$-factors of a pure HH state are $g_\parallel=3q=0.18$ and $g_\perp=6\kappa+27q/2=21.27$, while those of a pure LH state are $g_\parallel\approx 4\kappa=13.64$ and $g_\perp\approx 2\kappa=6.82$ (with $\kappa=3.41$ and $q=0.06$ the isotropic and cubic Zeeman parameters of Ge). The confinement and magnetic vector potential admix LH components into the HH ground-state; to lowest order in perturbation, the $g$-factors of the dot then read \cite{Michal21,martinez2022hole,Abadillo2023} 
\begin{subequations}
\label{eq:gHH}
\begin{align}
g_\parallel&\approx 3q+\frac{6}{m_0\Delta_\mathrm{LH}}(\lambda\langle p_x^2\rangle-\lambda^\prime\langle p_y^2\rangle) \\
g_\perp&\approx 6\kappa+\frac{27}{2}q-2\gamma_h\,,
\end{align}
\end{subequations}
where $\lambda=\kappa\gamma_2-2\eta_h\gamma_3^2$ and $\lambda^\prime=\kappa\gamma_2-2\eta_h\gamma_2\gamma_3$, with $\gamma_2=4.24$ and $\gamma_3=5.69$ the Luttinger parameters of bulk Ge \cite{Abadillo2023,Winkler03}. $\Delta_\mathrm{LH}$ is the HH/LH bandgap and $\langle p_x^2\rangle=\langle p_y^2\rangle\propto 1/\ell_\parallel^2$ are the expectation values of the squared in-plane momentum operators over the ground-state HH envelope. The factors $\gamma_h$ and $\eta_h$ depend on vertical confinement and describe the action of the magnetic vector potential on the orbital motion of the holes \cite{Ares13,Michal21}.  

For small enough vertical electric fields, the HH/LH bandgap of a Ge well can be approximated as \cite{Michal21}
\begin{equation}
\Delta_\mathrm{LH}\approx\frac{2\pi^2\hbar^2\gamma_2}{m_0L_w^2}+2b_v(\varepsilon_\parallel-\varepsilon_\perp)\,,
\label{eq:delta}
\end{equation}
where the first term accounts for vertical confinement \footnote{This expression is not valid for large $L_w$'s where the confinement becomes electrical and $\Delta_\mathrm{LH}\propto 1/\ell_\perp^2$. We use it as a guide in the present, qualitative discussion.} and the second term accounts for biaxial strains, with $b_v=-2.16$\,eV the uniaxial deformation potential of the valence band. In all but the thinnest strained Ge wells, the contribution from strains ($46$\,meV) overcomes confinement ($\approx 25$\,meV at $L_w=15$\,nm). Owing to the large $\Delta_\mathrm{LH}$ and dot sizes, $g_\parallel\approx 3q$; $g_\perp$ decreases when thinning the well due to the dependence of $\gamma_h$ on the vertical confinement profile \cite{Michal21}. 

In unstrained Ge wells, the HH/LH bandgap is ruled by vertical confinement only so that the HH/LH mixing is much stronger. This is highlighted by Fig.~\ref{fig:gfactLw}, which plots the weight $m^2$ of the LH components in the ground-state as a function of $L_w$. The mixing is $<0.2\%$ in strained Ge wells and decreases because $\ell_\parallel^2$ increases faster than $1/\Delta_\mathrm{LH}$. On the opposite, $m^2$ increases continuously in unstrained Ge wells and reaches $\approx 17.7\%$ in the bulk device. It remains nevertheless weak (and within the reach of the above perturbation theory) for thin $L_w$'s. The effect on the out-of-plane $g$-factor is impressive, $g_\perp$ being as small as $1.13$ in the bulk Ge device. This is actually much smaller than expected for a pure light-hole due to the complex interplay between the confinement and the magnetic spin and orbital Hamiltonians. 

The in-plane $g$-factor remains small and shows a non-monotonic behavior with increasing $L_w$. The change of sign of $g_\parallel$ around $L_w=25$\,nm is consistent with the closing of the HH/LH bandgap, thus the increase of the $\propto(\lambda-\lambda^\prime)<0$ correction in Eq.~\eqref{eq:gHH} \cite{martinez2022hole,Abadillo2023}. The bounce at large $L_w$ is due to the higher order terms not captured by this equation. At very large mixing, $g_\parallel$ shall tend to the in-plane $g$-factor $4\kappa=13.64$ of a pure light-hole.

\begin{figure}[t]
\centering
\includegraphics[width=\linewidth]{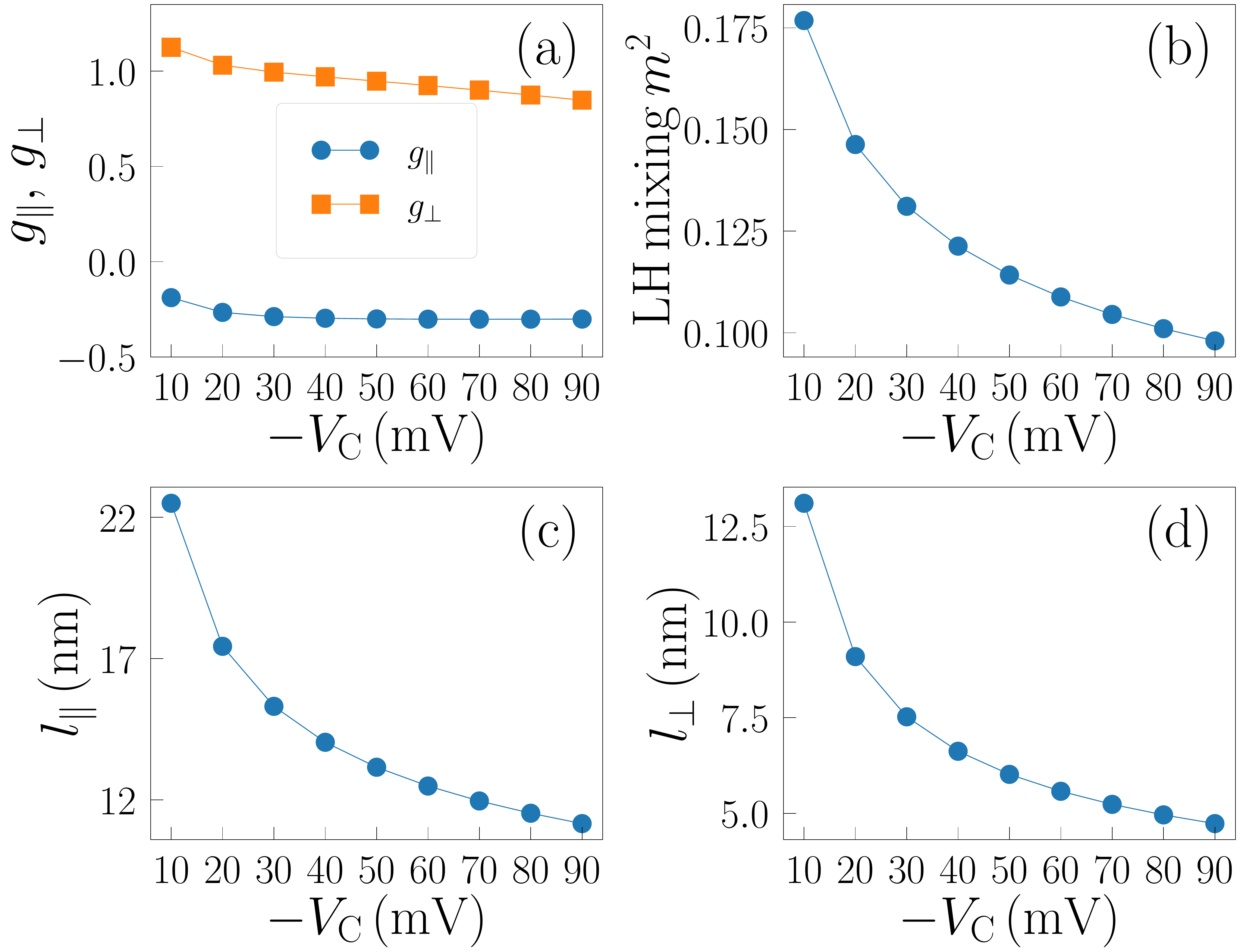}
\caption{(a) $g$-factors $g_\parallel$ and $g_\perp$, (b) LH mixing $m^2$, (c-d) extensions $l_\parallel$ and $l_\perp$ of the dot as as a function of $V_\mathrm{C}$ in the unstrained, bulk germanium device.} 
\label{fig:Vc}
\end{figure}

We finally discuss the dependence of the $g$-factors on the gate voltage $V_\mathrm{C}$ in the unstrained, bulk device. The dot extensions, $g$-factors, and HH/LH mixing in this device are plotted as a function of $V_\mathrm{C}$ in Fig.~\ref{fig:Vc}. As expected, the dot shrinks when $V_\mathrm{C}$ is pulled down, because the vertical and lateral components of the electric field are both $\propto V_\mathrm{C}$. As a consequence, the HH/LH bandgap opens, but the $\propto\langle p^2\rangle$ strength of the HH/LH couplings increases, so that the $g$-factors decrease rather slowly. The in-plane $g$-factor saturates to $g_\parallel\approx -0.3$ at high electric field.

\subsection{Spin manipulation}

\begin{figure*}[t]
\centering
\includegraphics[width=0.8\linewidth]{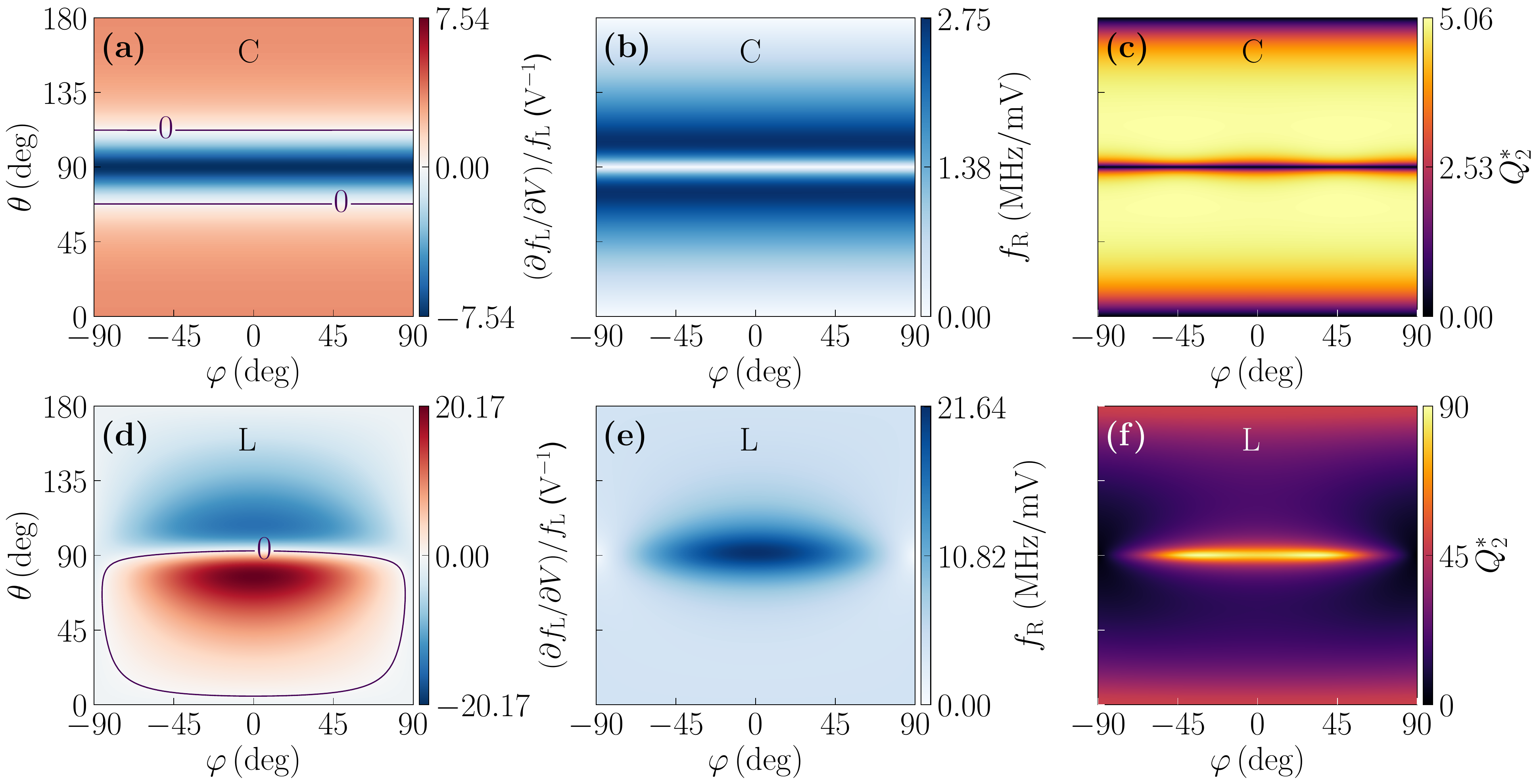}
\caption{Spin manipulation metrics in the unstrained, bulk Ge device. (a-c) Normalized LSES $(\partial f_\mathrm{L}/\partial V)/f_\mathrm{L}$, Rabi frequency $f_\mathrm{R}/V_\mathrm{ac}$ at constant Larmor frequency $f_\mathrm{L}=1$\,GHz, and quality factor $Q_2^*$ of the C gate as a function of the orientation of the magnetic field. (d-f) Same for L gate. The bias voltage is $V_\mathrm{C}=-25$\,meV.} 
\label{fig:Rabi}
\end{figure*}

\begin{figure*}[t]
\centering
\includegraphics[width=0.8\linewidth]{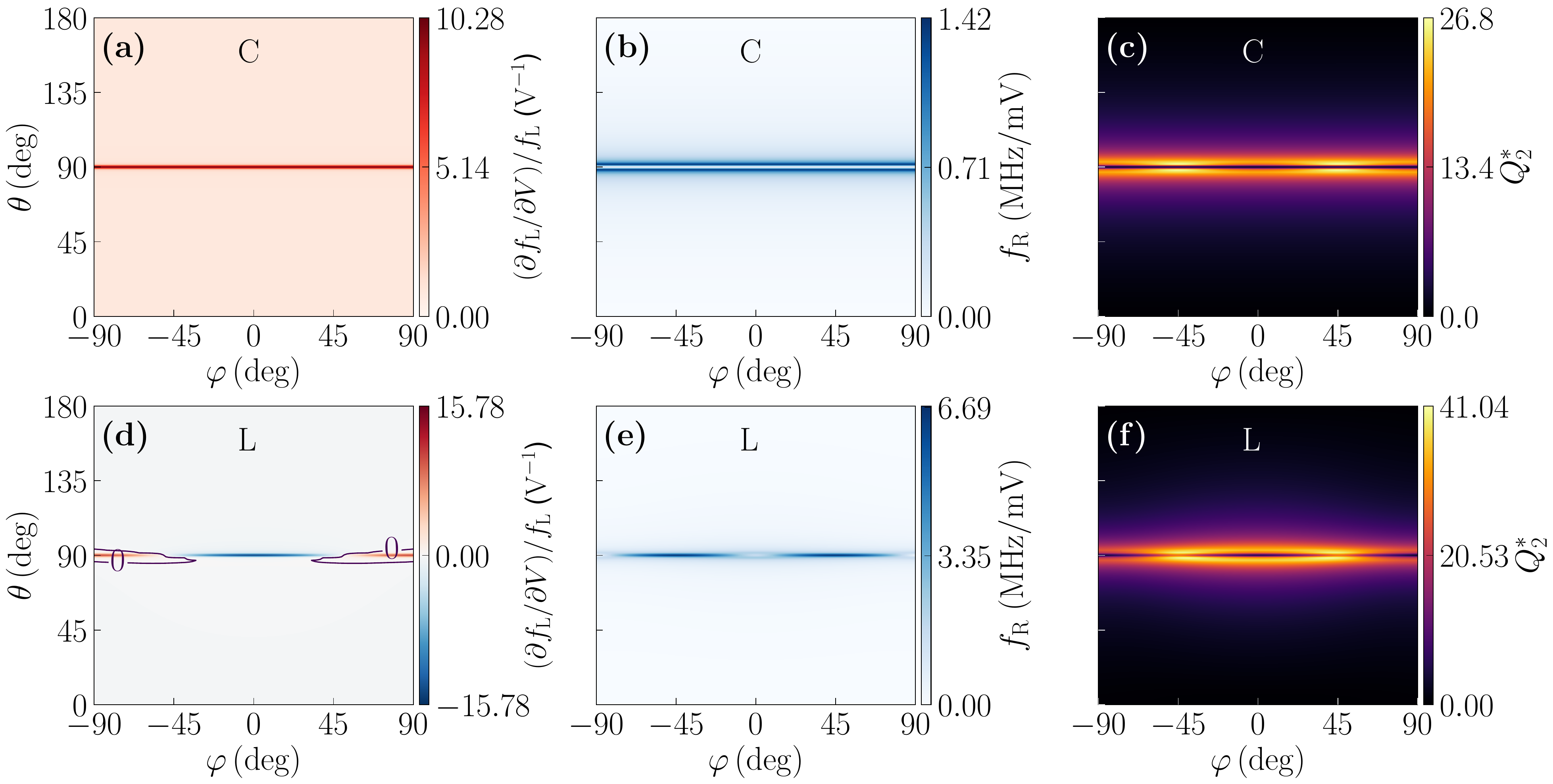}
\caption{Spin manipulation metrics in a strained Ge well with thickness $L_w=16$\,nm. (a-c) Normalized LSES $(\partial f_\mathrm{L}/\partial V)/f_\mathrm{L}$, Rabi frequency $f_\mathrm{R}/V_\mathrm{ac}$ at constant Larmor frequency $f_\mathrm{L}=1$\,GHz, and quality factor $Q_2^*$ of the C gate as a function of the orientation of the magnetic field. (d-f) Same for L gate. The bias voltage is $V_\mathrm{C}=-25$\,meV.} 
\label{fig:Rabi16}
\end{figure*}

We characterize spin manipulation with a given gate by the Rabi frequency $f_\mathrm{R}$ and by the quality factor $Q_2^*=2f_\mathrm{R}T_2^*$ (the number of $\pi$ rotations that can be achieved within the electrical dephasing time $T_2^*$). The Rabi frequency is computed from the $g$-matrix as \cite{Venitucci18}
\begin{equation}
f_\mathrm{R}=\frac{\mu_B|\vec{B}|V_\mathrm{ac}}{2hg^*}\left|(\hat{g}\mathbf{b})\times(\hat{g}^\prime\mathbf{b})\right|\,,
\label{eq:fR}
\end{equation}
where $\vec{b}=\vec{B}/|\vec{B}|$, $g^*=|\hat{g}\vec{b}|$ is the effective $g$-factor, $\hat{g}^\prime$ is the derivative of $\hat{g}$ with respect to the driving gate voltage, and  $V_\mathrm{ac}$ is the amplitude of the drive. To compute $T_2^*$, we lump as in Refs.~\cite{Mauro24,Mauro25} all electrical fluctuations into effective gate voltage noises:
\begin{equation}
\frac{1}{T_2^*}=\Gamma_2^*=\sqrt{2}\pi\sqrt{\sum_\mathrm{G}\left(\delta V_\mathrm{rms}\frac{\partial f_\mathrm{L}}{\partial V_\mathrm{G}}\right)^2}\,.
\label{eq:gammatot}
\end{equation}
The sum runs over the gates $\mathrm{G}\in\{\mathrm{C},\mathrm{L},\mathrm{R},\mathrm{T},\mathrm{B}\}$, $f_\mathrm{L}=\mu_B g^*|\mathbf{B}|$ is the Larmor frequency, $\delta V_\mathrm{rms}$ is the rms amplitude of the noise (assumed the same on all gates), and $\partial f_\mathrm{L}/\partial V_\mathrm{G}$ is the longitudinal spin electric susceptibility (LSES) of gate G (also a function of the corresponding $\hat{g}^\prime$) \cite{Piot22,Mauro24,Bassi24}. $Q_2^*$ is independent on $B=|\vec{B}|$ and is proportional to the ratio $\rho_\mathrm{ac}=V_\mathrm{ac}/\delta V_\mathrm{rms}$ between the drive and noise amplitudes. We set $\rho_\mathrm{ac}=100$ in the following. We discuss the hyperfine dephasing rate, as well as the spin-phonon relaxation time $T_1$ (which does not limit the operation of the qubits at $f_\mathrm{L}\lesssim 1$\,GHz) in the Supplementary Information.

The maps of the LSES, Rabi frequency and quality factor of the C and L gates are shown in Fig.~\ref{fig:Rabi} for the unstrained, bulk Ge device. The gate voltage is $V_\mathrm{C}=-25$\,mV and the principal $g$-factors are $g_\parallel=-0.28$ and $g_\perp=1.01$. The Rabi frequencies are plotted at constant Larmor frequency $f_\mathrm{L}=1$\,GHz. The maps of the B, R and T gates can be deduced from those of the L gate by rotations $\delta\varphi=90,\,180$, and $270^\circ$, respectively. The C gate modulates $\ell_\parallel$ and $\ell_\perp$ but does not break the symmetry of the dot, which only changes the diagonal $g$-factors $g_\parallel$ and $g_\perp$ ($g_\parallel^\prime=2.11$\,V$^{-1}$ and $g_\perp^\prime=3.5$\,V$^{-1}$). This has no effect on the spin precession axis when the magnetic field goes in-plane because the effective $g$-factors $|g_{xx}|$ and $|g_{yy}|$ remain degenerate ($f_\mathrm{R}\to0$ when $\theta\to90^\circ$). The Rabi frequency thus peaks out-of-plane but the C gate is far less efficient than the L gate for almost any magnetic field orientation. The $g^\prime$ matrix of the L gate reads:
\begin{align}
\hat{g}_\mathrm{L}^\prime=
\begin{pmatrix}
-1.66 & 0 & 8.40 \\
0  & -0.60 &  0 \\
12.02  & 0 & -0.88
\end{pmatrix}\,\mathrm{V}^{-1}\,.
\label{eq:GLprime}
\end{align}
This matrix is dominated by the $g^\prime_{xz}$ and $g^\prime_{zx}$ terms that capture the rotations of the principal axes of the $g$-matrix in the inhomogeneous electric field of the L gate \cite{martinez2022hole} (and also the effects of a cubic Rashba interaction \cite{Marcellina17,Terrazos21,Bosco21b}). Namely, the axis of $z'$ of the $g_\perp$-factor (and the orthogonal, equatorial $(x'y')$ plane) rock from left to right when the dot is driven by the L gate, which tilts the precession axis of the spin and results in coherent spin rotations at resonance. This mechanism gives rise to the prominent peak for in-plane magnetic fields, because the effects of small excursions of $\vec{B}$ around the effective equatorial $(x'y')$ plane are amplified by $g_\perp>g_\parallel$, and because the Rabi frequency is $\propto B$, which is larger in-plane at constant Larmor frequency.

The LSES of the C gate is maximal in-plane but displays two ``sweet'' lines (zero LSES) at $\theta=90\pm 22^\circ$ \cite{Mauro24}. Likewise, the LSESs of the side gates (which primarily characterize the sensitivity to lateral electric field noise) show sweet lines running near the equatorial plane. Along these sweet lines, the hole decouples (to first-order) from the corresponding component of the noise. The Rabi frequency of a given gate reaches its maximum near the sweet line(s) of that gate owing to ``reciprocal sweetness'' arguments \cite{michal2022tunable,Bassi24}. As the C gate is inefficient in-plane, $Q_2^*(\mathrm{C})$ broadly peaks for $\theta=90\pm 22^\circ$, while $Q_2^*(\mathrm{L})$ peaks in-plane as does the Rabi frequency. The quality factors achieved with the L gate are, nevertheless, much larger since the Rabi oscillations are faster. 

The same maps are plotted in a reference, strained Ge well with thickness $L_w=16$ nm in Fig.~\ref{fig:Rabi16}. The Rabi map is qualitatively similar to Fig.~\ref{fig:Rabi} for the C gate, but the LSES has no sweet lines since all diagonal elements of $g_\mathrm{C}^\prime$ have the same sign \cite{Mauro24}. The Rabi oscillations (and LSES) of the L gate are now dominated by the modulations of the diagonal $g$-factors because the vertically more confined wave function does not ``rock'' as much  \cite{martinez2022hole}. As a consequence, the Rabi frequency still peaks in-plane but near $\varphi=\pm 45^\circ$. All features are, however, much thinner (and closer to the equatorial plane) in the strained well than in the bulk device. Indeed, the spin precession axis gets locked onto $z$ once the magnetic field goes slightly out-of-plane in strained Ge owing to the large $g_\perp/g_\parallel$ ratio \cite{Mauro24}. For the L gate, the full width at half-maximum (FWHM) of the Rabi peak on Fig.~\ref{fig:Rabi}e is $\delta\theta=39^\circ$ and the FWHM of the $Q_2^*$ peak on Fig.~\ref{fig:Rabi}f is $\delta\theta=12.3^\circ$ ({\it vs} $1.7^\circ$ and $4.8^\circ$, respectively, on Fig.~\ref{fig:Rabi16}e-f). Contrary to the strained Ge well, the sweet lines of the C and L gates are well separated from the hot spots (maximal LSES) where the sensitivity to noise is enhanced. 

Another striking difference between the bulk device and the reference, strained Ge well is the magnitude of the Rabi frequencies. Although the balance between the driving mechanisms is not the same, the effects of SOC are generally expected to be enhanced by a reduction of the HH/LH bandgap. For the L gate, the maximal Rabi frequency is $f_\mathrm{R}/V_\mathrm{ac}=21.6$\,MHz/mV in the bulk device but $f_\mathrm{R}/V_\mathrm{ac}=6.7$\,MHz/mV in the strained Ge well. The LSES, thus the sensitivity to noise is also enhanced, yet $Q_2^*$ is significantly larger (and broader) in the bulk device.

\begin{figure}[t]
\centering
\includegraphics[width=\linewidth]{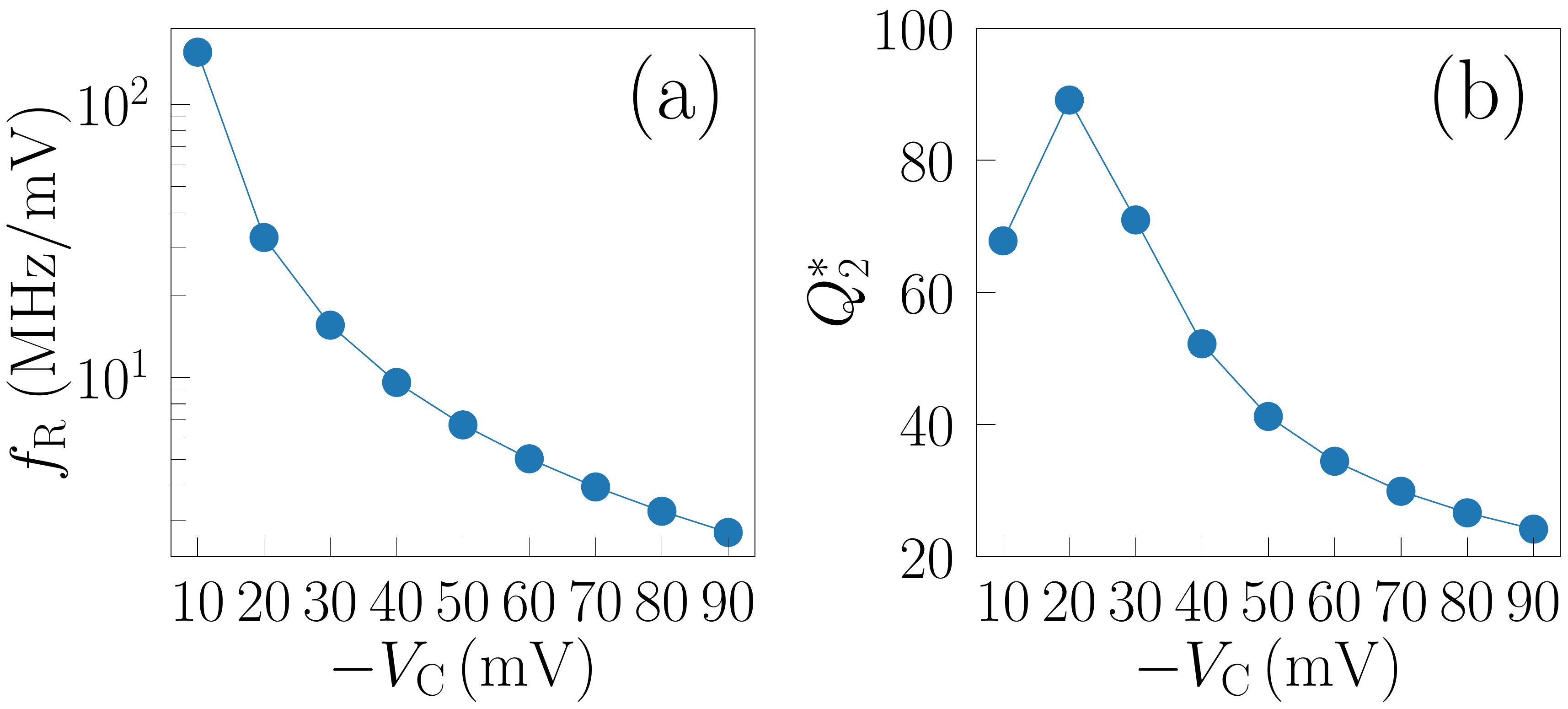}
\caption{(a) Rabi frequency and (b) quality factor of the L gate of the unstrained, bulk Ge device as a function of $V_\mathrm{C}$, for a magnetic field $\vec{B}\parallel\vec{x}$.} 
\label{fig:QVc}
\end{figure}

Finally, we plot in Fig.~\ref{fig:QVc} the Rabi frequency and quality factor of the L gate of the bulk Ge device as a function of $V_\mathrm{C}$, for a magnetic field $\vec{B}\parallel\vec{x}$. The Rabi frequency decreases as $V_\mathrm{C}$ is pulled down because the dot gets smaller (see Fig.~\ref{fig:Vc}), thus less responsive to the drive field (and the HH/LH bandgap opens). The quality factor also decreases, but the optimal magnetic field orientation (best $Q_2^*$) moves towards $\varphi=\pm 45^\circ$. For $\vec{B}\parallel\vec{x}$, the quality factor is optimal for $V_\mathrm{C}\approx -25$\,mV.

\section{Discussion}

The above calculations illustrate the benefits in making hole spin qubits in unstrained Germanium. Without biaxial strains, the HH/LH bandgap closes and the HH/LH mixing is enhanced. The strong anisotropy of the gyromagnetic response, characteristic of pure heavy-holes, is therefore significantly reduced. The ratio $g_\perp/g_\parallel$ in the ground-state thus decreases from $\gtrsim 50$ in strained heterostructures to $\approx 3$ in unstrained, bulk Ge devices (which is comparable to some silicon hole spin qubits \cite{Piot22,Bassi24}). This broadens the features in the maps of Rabi frequency and quality factor, thus extends the operational range of the devices, which shall ease the scaling to many qubits. This softer gyromagnetic anisotropy is one of the strongest assets of the bulk Ge route.
As another advantage, the density of dislocations shall be much lower in bulk Ge than in strained heterostructures grown on GeSi buffers.

The larger HH/LH mixing generally enhances the effects of SOC. As a consequence, the Rabi oscillations are significantly faster than in strained heterostructures, but the electrical dephasing times are shorter (for a given level of noise). Nonetheless, the quality factors for single spin manipulation can be larger in bulk Ge than in strained heterostructures -- namely, the Rabi frequencies increase faster than the dephasing rate in a broad range of magnetic field orientations. However, this is practically advantageous only if the two-qubit gates are not much slower than the single-qubit operations. 

In order to limit the degradation of $T_2^*$, the level of charge disorder and noise must be carefully controlled in bulk Ge devices, especially because the top GeSi barrier will typically be thinner ($\approx 20$\,nm) than in strained heterostructures to avoid plastic relaxation. The materials and layout of the gate stack must, therefore, be engineered to limit the density of charged defects \cite{Varley23,Massai23,Martinez24}. The hyperfine dephasing times \cite{Bosco21c} are comparable or even better in bulk Ge devices, except for strictly in-plane magnetic fields (see Supplementary Information). All germanium spin qubits would, nevertheless, strongly benefit from isotopic purification. The operation of bulk Ge qubits will likely be optimal at small Larmor frequencies $f_\mathrm{L}$ \cite{Valentin25} where the single qubit operations are much faster than in strained heterostructures but the electrical $T_2^*\propto 1/B$ remains long enough.

Despite the enhanced HH/LH mixing, the ground-state still exhibits a dominant HH character (LH mixing $10$ to $20\%$). The first and higher excited orbitals (relevant for many-holes qubits \cite{Valentin25}) may, however, show much larger (even prevailing) LH components. While probing the physics of these highly mixed states is certainly interesting, reliable and reproducible results may call for a tight control of the dot occupations.

To conclude, we point out that the strength of the HH/LH mixing can be finely tuned by decoupling the compositions of the GeSi buffer and top barrier. The Ge well may indeed be grown on a thick GeSi buffer with low Si fraction, and capped with a GeSi layer with larger concentration (to achieve a robust barrier). The small compressive strains imposed by the GeSi buffer will slightly open the HH/LH bandgap and mitigate the effects of SOC (at the price of a larger $g$-factor anisotropy). They will also allow for a thicker GeSi barrier. The optimal buffer concentration results from a compromise between the target dephasing time $T_2^*$ and $g$-factor anisotropy, thus depends on the level of noise and variability. 

\vspace{1.5cm}
{\bf ACKNOWLEDGEMENTS} \\

This work was supported by the ``France 2030'' program (PEPR PRESQUILE-ANR-22-PETQ-0002) and by the Horizon Europe Framework Program (grant agreement 101174557 QLSI2).

\setcounter{section}{0}
\setcounter{equation}{0}
\setcounter{figure}{0}
\setcounter{table}{0}

\renewcommand\thefigure{S\arabic{figure}} 
\renewcommand\theequation{S\arabic{equation}}

\vspace{1cm}

\clearpage

\begin{widetext}
\begin{center}
\textbf{\large Supplementary Information for ``Hole spin qubits in unstrained Germanium layers''}
\end{center}
\end{widetext}

In this supplementary information, we discuss the phonon-limited relaxation rates, the dephasing by hyperfine interactions, and the effects of the inhomogeneous strains induced by the thermal contraction of materials when the device is cooled down.

\section{Relaxation rates by hole-phonon interactions}

We compute the phonon-limited relaxation rates of the devices with the methodology of Ref.~\cite{Li20}. We assume bulk-like acoustic phonons coupled to the holes through the valence band deformation potentials $a_v=2$\,eV, $b_v=-2.16$\,eV and $d_v=-6.06$\,eV \cite{Abadillo2023}. The longitudinal and transverse sound velocities are $v_l=5300$\,m/s and $v_t=3300$\,m/s.

\begin{figure}
\centering
\includegraphics[width=.8\linewidth]{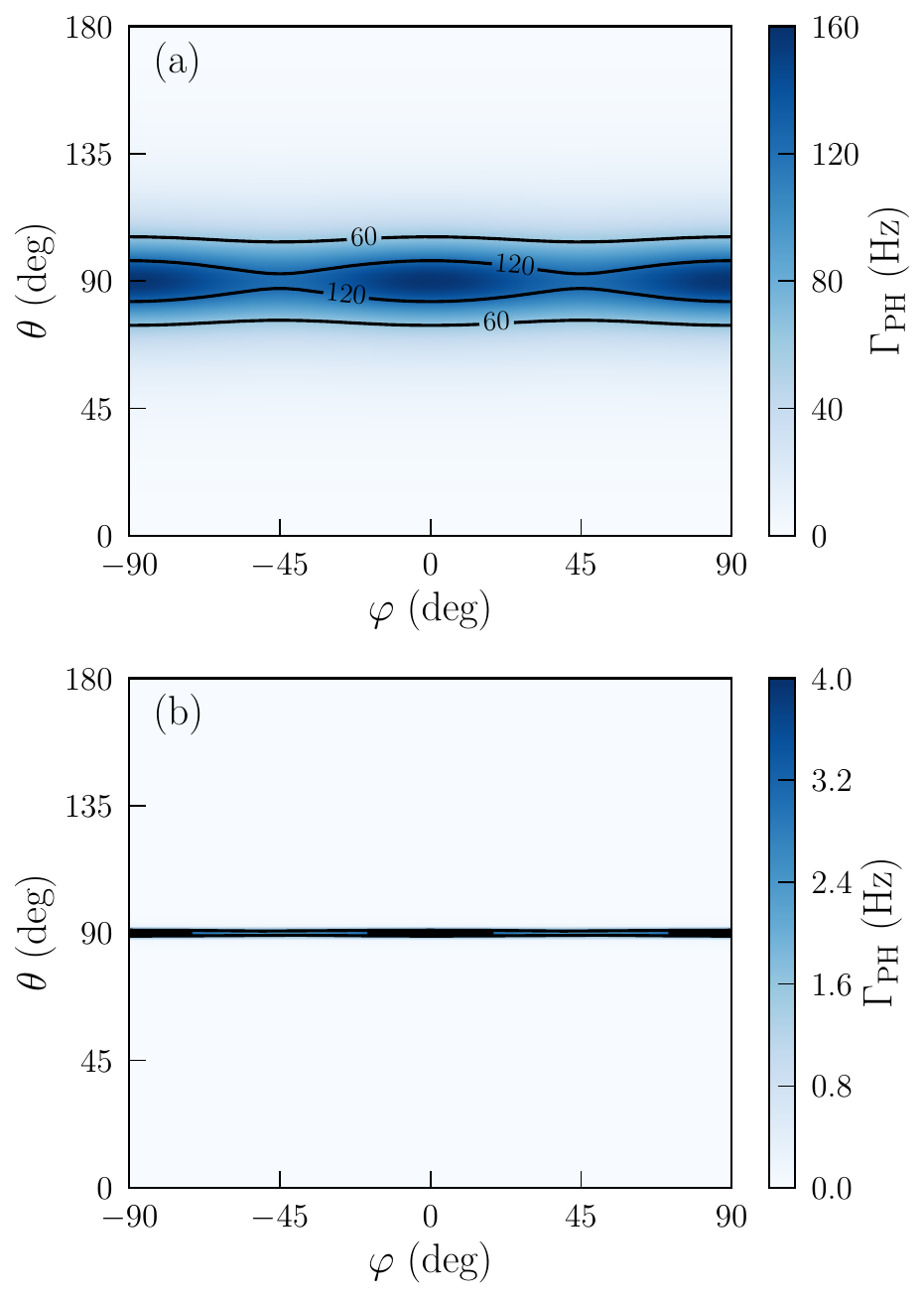}
\caption{Phonon-limited relaxation rate as a function of the magnetic field orientation in (a) the unstrained, bulk Ge device and (b) the reference, strained heterostructure with $L_w=16$\,nm. The bias voltage is $V_\mathrm{C}=-25$\,mV, the Larmor frequency $f_\mathrm{L}=1$\,GHz, and the temperature $T=100$\,mK.} 
\label{fig:gammaph}
\end{figure}

The relaxation rates $\Gamma_\mathrm{PH}$ of the unstrained, bulk Ge device and of the reference, strained heterostructure with $L_w=16$\,nm are plotted as a function of the magnetic field orientation in Fig.~\ref{fig:gammaph}. The bias voltage is $V_\mathrm{C}=-25$\,mV, the Larmor frequency is $f_\mathrm{L}=1$\,GHz, and the temperature is $T=100$\,mK.

The relaxation rate peaks for in-plane magnetic fields. It is dominated by the effects of the shear strains induced by the phonons (thus by the deformation potential $d_v$). As a consequence, $\Gamma_\mathrm{PH}$ scales $f_\mathrm{L}^4$ when $hf_\mathrm{L}\ll kT$ and as $f_\mathrm{L}^5$ when $hf_\mathrm{L}\gg kT$ \cite{Li20}. Due to the smaller HH/LH bandgap and stronger spin-orbit coupling effects, the relaxation rate is up to 40 times larger in the bulk Ge device than in the reference, strained heterostructure. The relaxation time $T_1=\Gamma_\mathrm{PH}^{-1}>6.5$\,ms remains, nonetheless, much larger than the usual dephasing times \cite{Valentin25} and should not, therefore, be limiting despite this enhancement (at least when $f_\mathrm{L}\lesssim 1$\,GHz). Note that the in-plane ``hot spot'' is much broader in the bulk Ge device than in the strained heterostructure owing (as for the Rabi frequency) to the reduction of the gyromagnetic anisotropy.

\section{Dephasing due to hyperfine interactions}

We compute the hyperfine dephasing time $T_{2,\mathrm{h}}^*$ with the methodology of Ref.~\cite{Piot22}. The hyperfine interactions between the hole and the $N$ nuclei spins are described by the Hamiltonian \cite{Machnikowski19,Bosco21c}
\begin{equation}
H_{\rm int}=\frac{A}{2n_0}\sum_{n=1}^N\, \delta(\mathbf{r}-\mathbf{R}_n)\otimes\mathbf{J}\cdot\mathbf{I}_n\,,
\end{equation}
where $A$ is the hyperfine coupling constant, $n_0$ is the density of nuclei, $\mathbf{I}_n$ is the spin operator of nucleus $n$ at position $\mathbf{R}_n$, and $\mathbf{J}$ is the angular momentum operator acting on the $J=3/2$ Bloch functions of the heavy and light holes. Assuming uncorrelated and unpolarized nuclear spins with Gaussian-distributed quasi-static fluctuations, the dephasing rate $\Gamma_{2,\mathrm{h}}^*=1/T_{2,\mathrm{h}}^*$ is then given by \cite{Merkulov02,fischer2008spin,testelin2009hole}
\begin{equation}
\Gamma_{2,\mathrm{h}}^*=\frac{|A|}{2\hbar}\sqrt{\frac{\nu I(I+1)}{6n_0}}\left(\overline{\delta J_x^2}+\overline{\delta J_y^2}+\overline{\delta J_z^2}\right)^{1/2}\,,
\end{equation}
where $\nu$ is the fraction of spin-carrying nuclei,
\begin{equation}
\overline{\delta J_\alpha^2}=\int d^3\mathbf{R}\,\delta J_\alpha^2\left(\mathbf{R}\right)\,,
\end{equation}
and
\begin{equation}
\delta J_\alpha(\mathbf{R})=\left\langle\uparrow\right| \delta(\mathbf{r}-\mathbf{R})\otimes J_\alpha\left|\uparrow\right\rangle-\left\langle\downarrow\right| \delta(\mathbf{r}-\mathbf{R})\otimes J_\alpha\left|\downarrow\right\rangle\,,
\end{equation}
with $\left|\uparrow\right\rangle$ and $\left|\downarrow\right\rangle$ the two Zeeman-split states at a given magnetic field orientation. $T_{2,\mathrm{h}}^*$ is independent on the Larmor frequency as the hyperfine interaction gives rise to a magnetic-like noise. On the opposite, electrical noise can couple to the spin only at finite magnetic field, so that the electrical dephasing rate $\Gamma_2^*$ [Eq. (4) of the main text] is $\propto B$ or $\propto f_\mathrm{L}$.

\begin{figure}
\centering
\includegraphics[width=.8\linewidth]{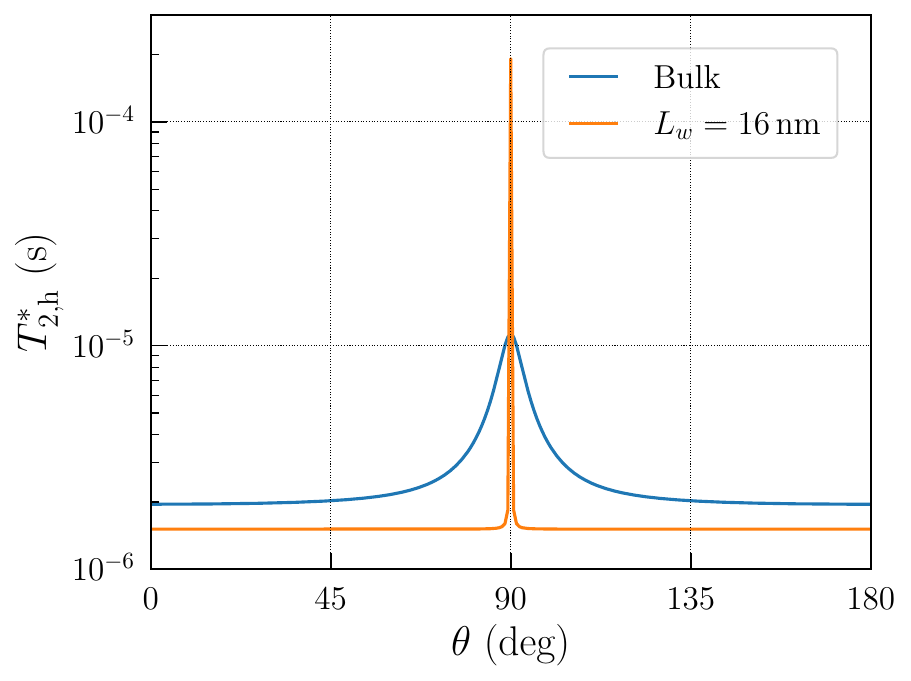}
\caption{Hyperfine dephasing rate $T_{2,\mathrm{h}}^*$ as a function of the magnetic field angle $\theta$ in the unstrained, bulk Ge device and in the reference, strained heterostructure with $L_w=16$\,nm. The bias voltage is $V_\mathrm{C}=-25$\,mV.} 
\label{fig:T2h}
\end{figure}

In germanium, we take $|A|=0.73$\,$\mu$eV, $I=9/2$, $n_0=44.17$\,nm$^{-3}$ and $\nu=0.077$ for the $^{73}$Ge atoms \cite{Bosco21}. We neglect dephasing by the $^{29}$Si atoms in Ge$_{0.8}$Si$_{0.2}$. The dephasing times $T_{2,\mathrm{h}}^*$ computed in the unstrained, bulk Ge device and in the reference, strained heterostructure with $L_w=16$\,nm are plotted as a function of the magnetic field angle $\theta$ in Fig.~\ref{fig:T2h}. They are weakly dependent on $\varphi$. 

As expected, the dephasing time of the mostly heavy-hole states peaks for in-plane magnetic fields (where pure HH $\left|\uparrow\right\rangle$ and $\left|\downarrow\right\rangle$ spins are mixed $J_z=\pm 3/2$ states with vanishing $\langle J_x\rangle$, $\langle J_y\rangle$ and $\langle J_z\rangle$). In the reference device, $T_{2,\mathrm{h}}^*$ decays very rapidly once the magnetic field goes out-of-plane and the spins become almost pure $J_z=\pm 3/2$ states. This dependence is much softened by the HH-LH mixing in the bulk Ge device. The in-plane $T_{2,\mathrm{h}}^*\approx 11$\,$\mu$s is, however, smaller. We emphasize, though, that an accurate description of the peak of the strained heterostructure goes beyond the present model (which misses, e.g., cubic anisotropy corrections \cite{Machnikowski19,Philippopoulos20}). For strongly out-of-plane magnetic fields, the hyperfine dephasing time is $1.5-2$\,$\mu$s in both devices, in agreement with recent experiments \cite{Hendrickx2024}. Bulk germanium devices as well as strained Ge/GeSi heterostructures would, therefore, benefit from isotopic purification.

\section{Effects of inhomogeneous strains}

In this section, we discuss the effects of the inhomogeneous strains imposed by the differential thermal contraction of materials.

Indeed, the metals contract much faster than the semiconductors when the device is cooled down to cryogenic temperatures. As a consequence, the gate stack can imprint small strains in the heterostructure. The resulting shear strains can, in particular, have a significant impact on the spin dynamics~\cite{Abadillo2023}.

We compute the strains in the bulk Ge device using a finite elements solver for the continuum elasticity equations. All material parameters (lattice constants, thermal contraction coefficients, ...) are taken from Ref.~\cite{Abadillo2023}. We consider 20-nm-thick aluminium gates matched to the Ge substrate at $T=300$\,K. The effective lattice mismatch between Al and Ge due to differential thermal contraction is thus $\varepsilon=-0.34\%$ at $T\to0$\,K. We assume that the whole structure relaxes elastically. In practice, grain boundaries in the metal, or plastic relaxation at the metal/oxide interface may limit the transfer of strains to the heterostructure. This approach has, nonetheless, proven successful in the analysis of the spin dynamics of donors in silicon strained by a metallic resonator \cite{Pla18,Ranjan21}.

\begin{figure*}
\centering
\includegraphics[width=0.3\linewidth]{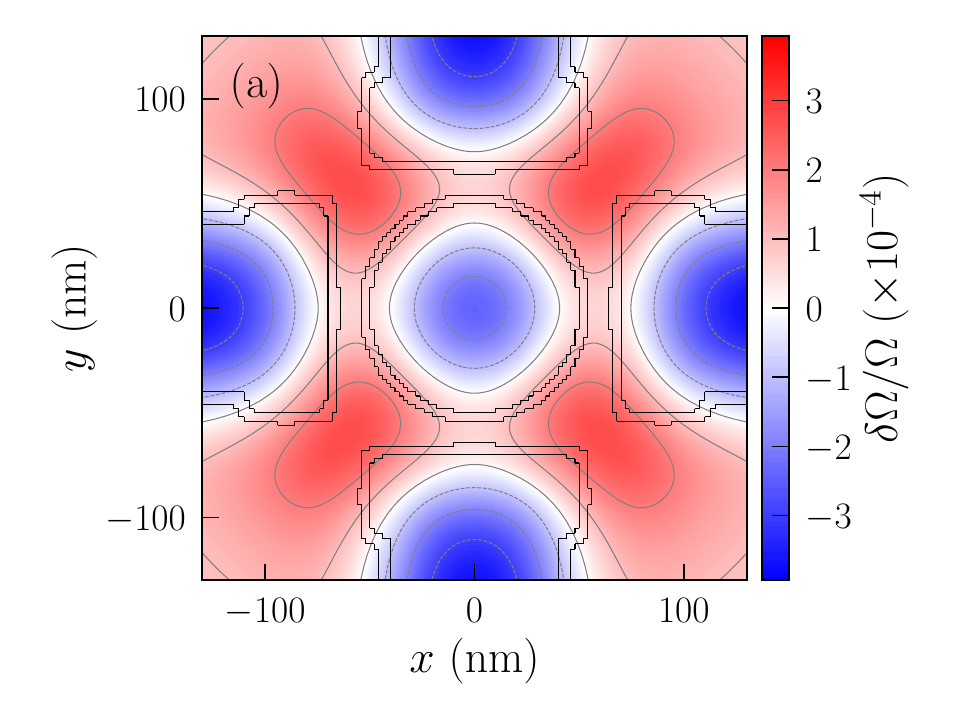}
\includegraphics[width=0.3\linewidth]{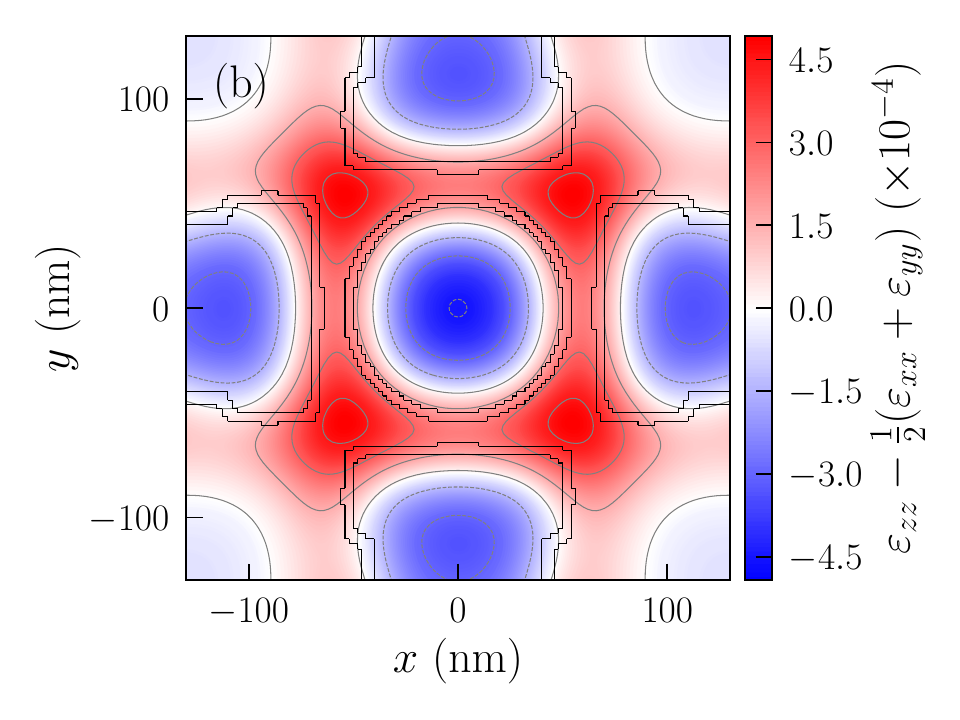}
\includegraphics[width=0.3\linewidth]{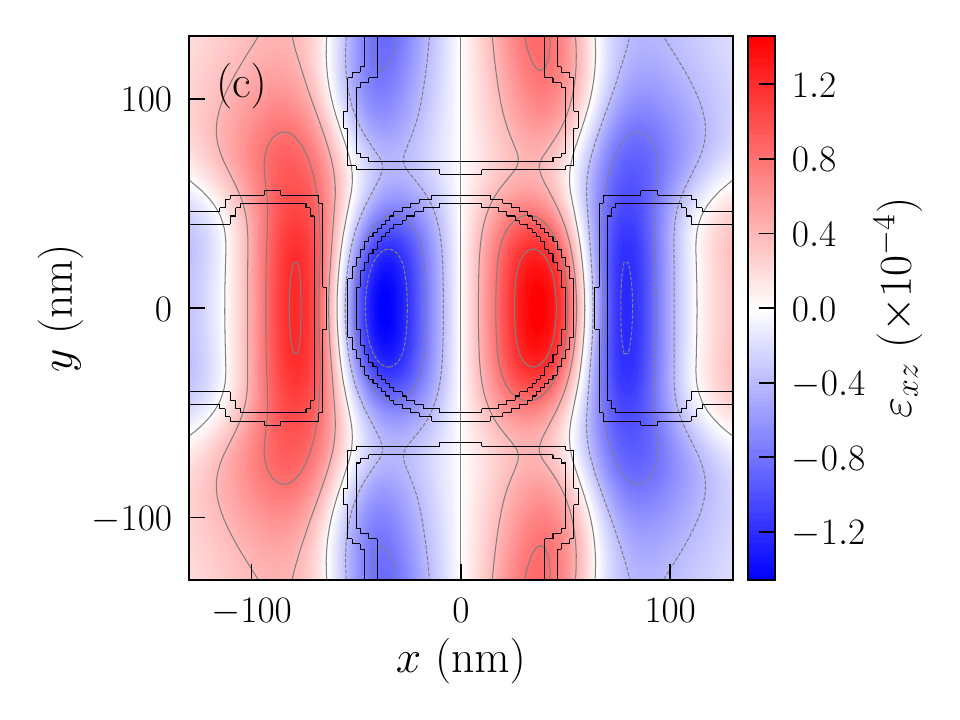}
\caption{Inhomogeneous strains in the bulk Ge device, in a $(xy)$ plane 10\,nm below the Ge/GeSi interface. The hydrostatic strain $\delta\Omega/\Omega=\varepsilon_{xx}+\varepsilon_{yy}+\varepsilon_{zz}$ is the local, relative variation of the volume of the material. The black lines delineate the position of the gates (and Al$_2$O$_3$ around) at the surface of the device.} 
\label{fig:strains}
\end{figure*}

Relevant strains are plotted in Fig.~\ref{fig:strains} in a horizontal $(xy)$ plane 10\,nm below the Ge/GeSi interface. The hydrostatic strain $\delta\Omega/\Omega=\varepsilon_{xx}+\varepsilon_{yy}+\varepsilon_{zz}$ is the local, relative change of the volume $\Omega$ of the unit cell of Ge, while $\varepsilon_\mathrm{uni}=\varepsilon_{zz}-(\varepsilon_{xx}+\varepsilon_{yy})/2$ characterizes the uniaxial component of the deformation, and $\varepsilon_{xz}$ the change of angle between the $x$ and $z$ axes of the unit cell. The inhomogeneous strains are rather small (in the few $10^{-4}$ range), and have the symmetry of the gate layout (except $\varepsilon_{xz}$, because a $90^\circ$ rotation transforms $\varepsilon_{xz}$ into $\varepsilon_{yz}$). As expected, the Ge substrate undergoes compressive strains below the gates ($\delta\Omega/\Omega<0$) balanced by tensile strains in-between. The uniaxial strains are also negative below the gates (but would have been positive if the stress imposed by the gates was biaxial). The shear strains are typically maximum near the transitions from compressive to tensile strains.

\begin{figure*}
\centering
\includegraphics[width=0.8\linewidth]{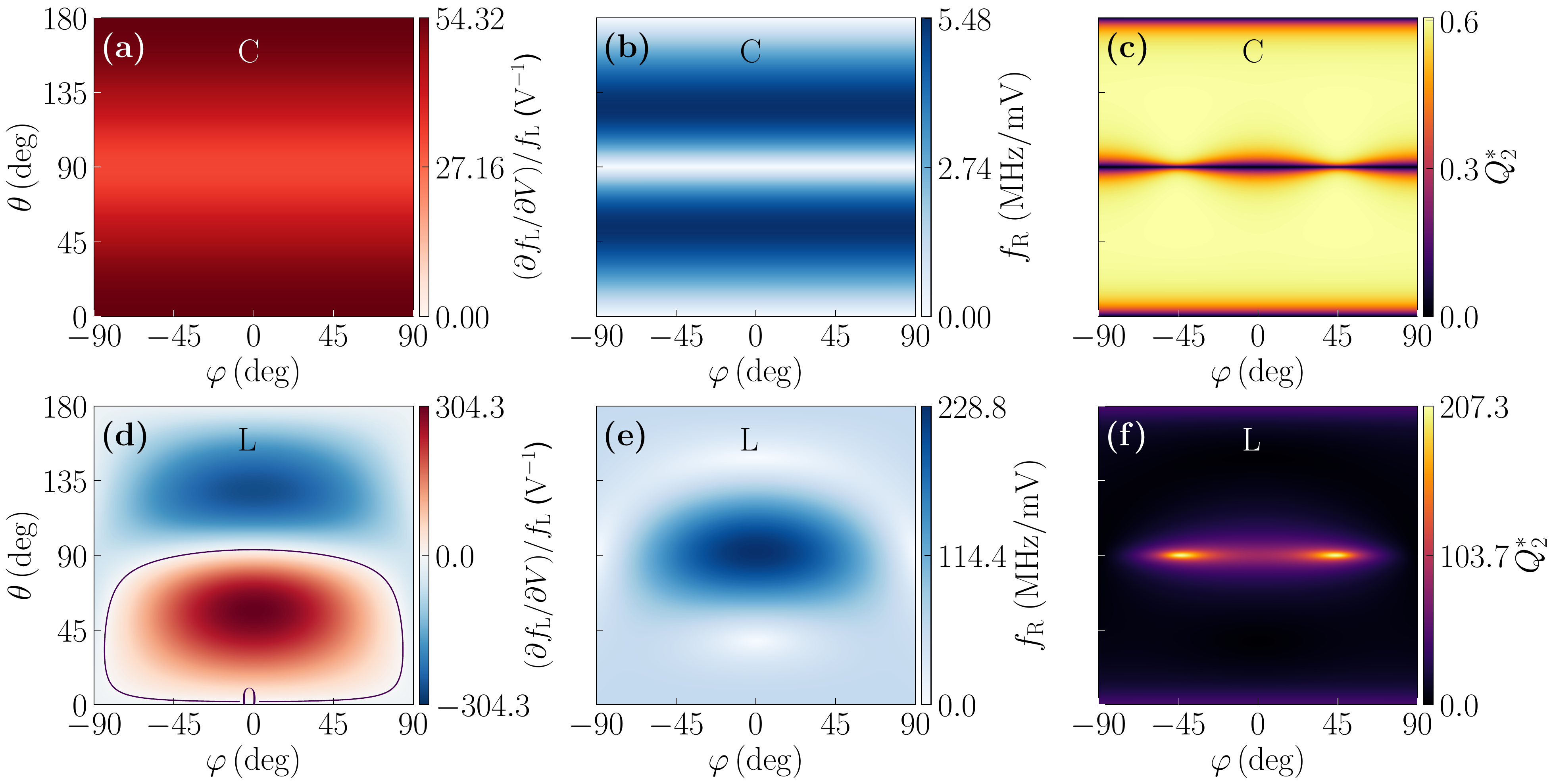}
\caption{Spin manipulation metrics in the bulk Ge device in the presence of inhomogeneous cool-down strains. (a-c) Normalized LSES $(\partial f_\mathrm{L}/\partial V)/f_\mathrm{L}$, Rabi frequency $f_\mathrm{R}/V_\mathrm{ac}$ at constant Larmor frequency $f_\mathrm{L}=1$\,GHz, and quality factor $Q_2^*$ of the C gate as a function of the orientation of the magnetic field. (d-f) Same for L gate. The bias voltage is $V_\mathrm{C}=-25$\,mV.} 
\label{fig:Rabi_strains}
\end{figure*}

\begin{figure}[!t]
\centering
\includegraphics[width=0.8\linewidth]{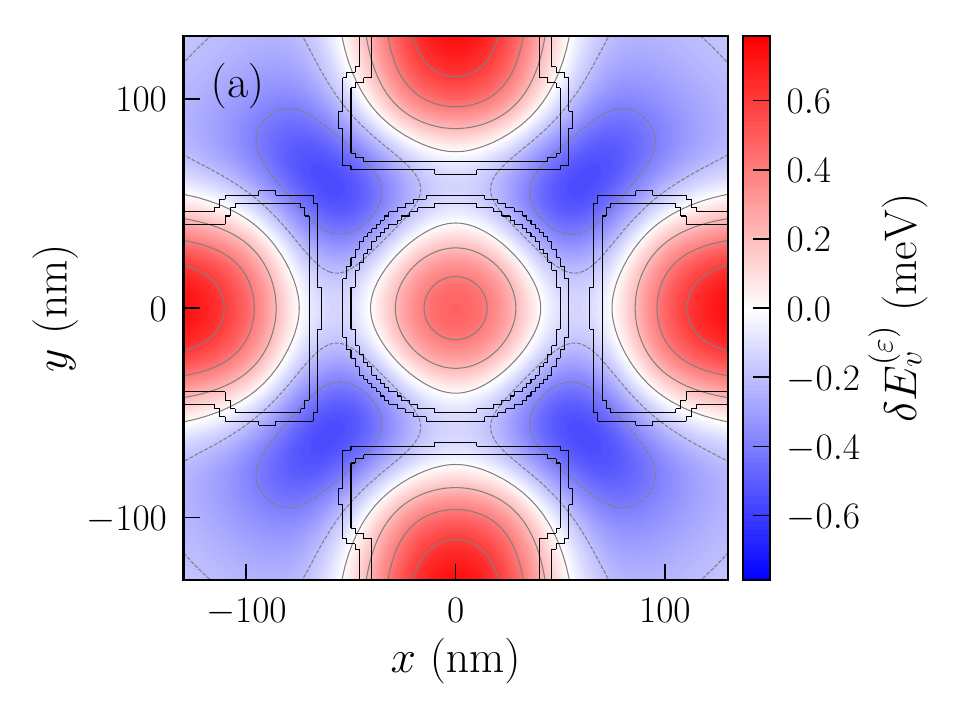}
\includegraphics[width=0.8\linewidth]{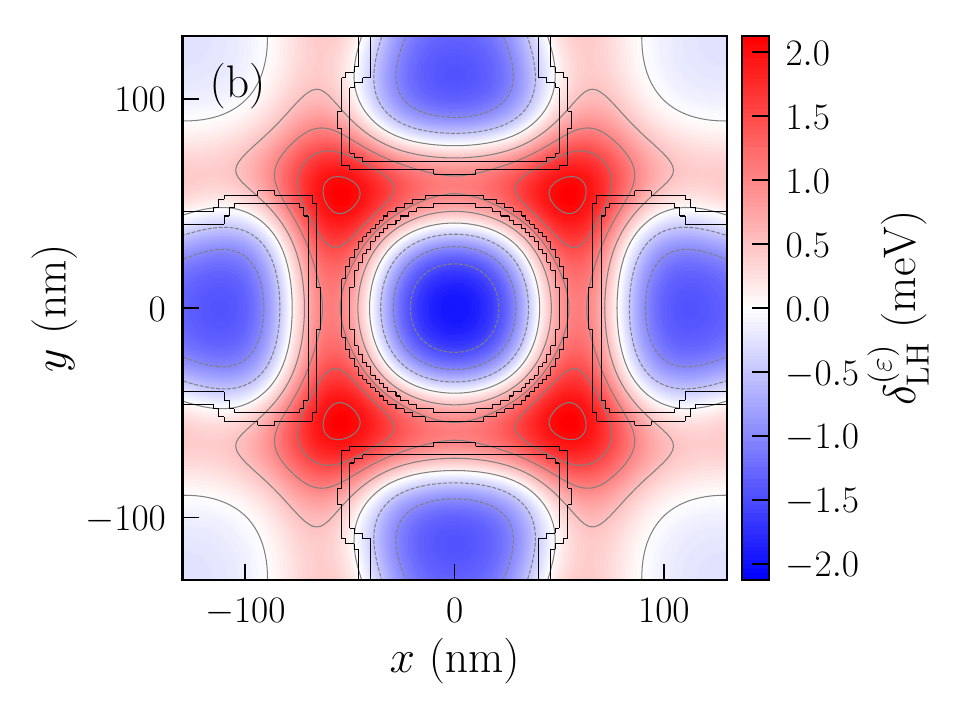}
\caption{Local variations of (a) the valence band edge $\delta E_v^{(\varepsilon)}$ and (b) the HH/LH bandgap $\delta_\mathrm{LH}^{(\varepsilon)}$ induced by inhomogeneous strains in the bulk Ge device, plotted in a $(xy)$ plane 10\,nm below the Ge/GeSi interface. The black lines delineate the position of the gates (and Al$_2$O$_3$ around) at the surface of the device.} 
\label{fig:edges}
\end{figure}

\begin{figure}[!t]
\centering
\includegraphics[width=0.8\linewidth]{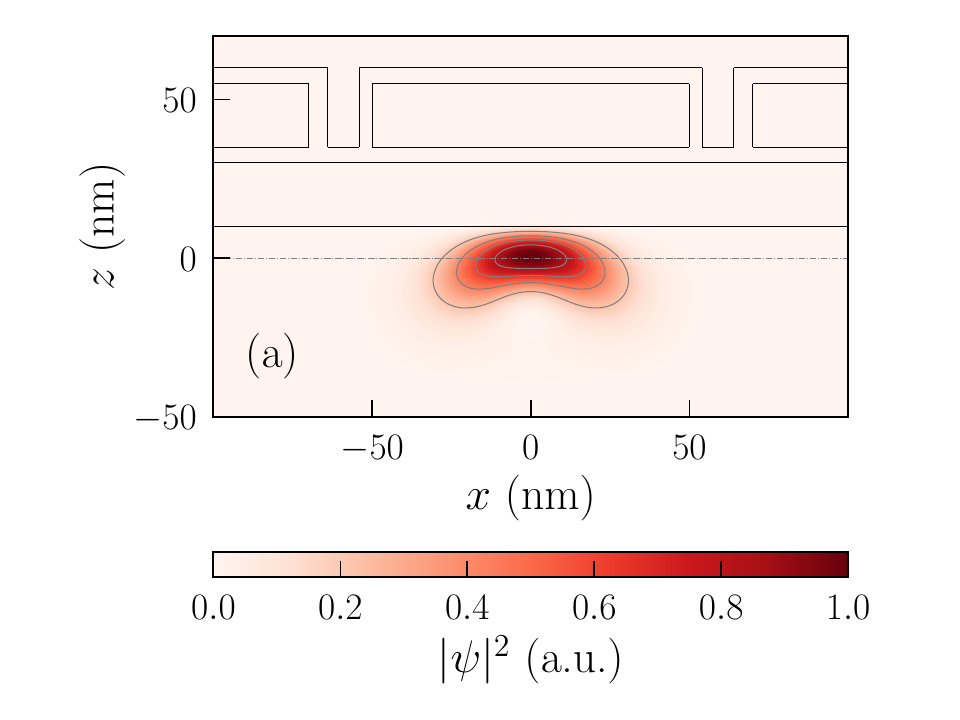}
\includegraphics[width=0.8\linewidth]{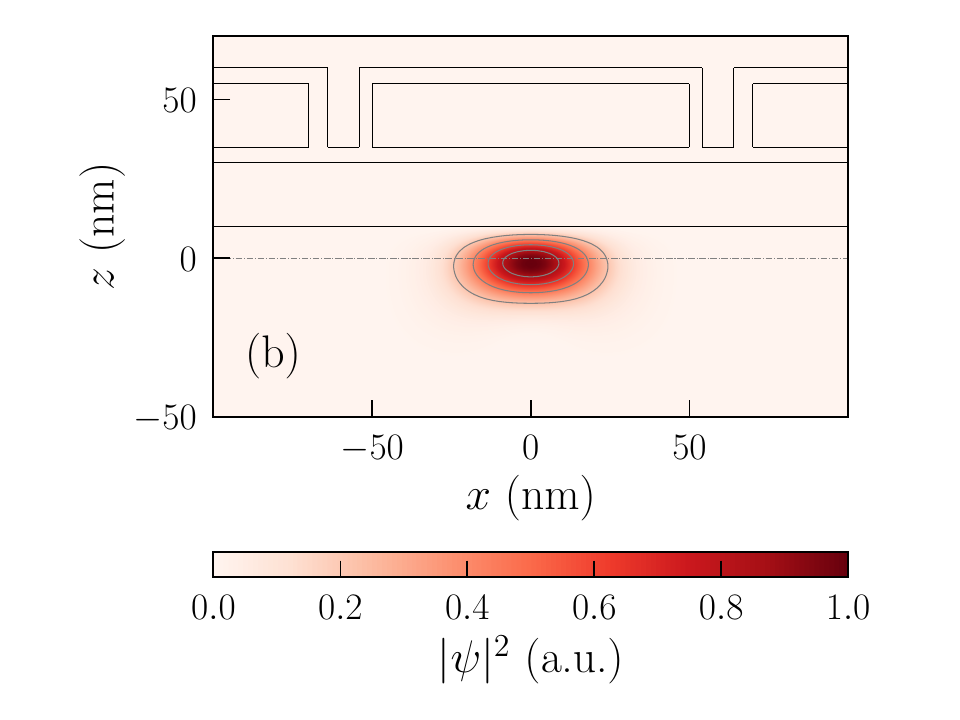}
\caption{The ground-state hole density $|\psi|^2$ in the bulk Ge device (a) with and (b) without inhomogeneous strains, plotted in a $(xz)$ plane at $y=0$ (the vertical symmetry plane of the device). The bias voltage is $V_\mathrm{C}=-25$\,mV. Figs.~\ref{fig:strains} and \ref{fig:edges} are plotted in the $(xy)$ plane at $z=0$ indicated by the dash-dotted line.} 
\label{fig:wfns}
\end{figure}

The maps of the LSES, Rabi frequency and quality factor of the C and L gates are plotted in Fig.~\ref{fig:Rabi_strains}. The gate voltage is $V_\mathrm{C}=-25$\,mV and the Rabi frequencies are computed at constant Larmor frequency $f_\mathrm{L}=1$\,GHz. These maps can be directly compared to Fig. 5 of the main text (where the bulk Ge substrate is unstrained and the GeSi barrier is homogeneously, biaxially strained). They are, overall, qualitatively similar (except for the LSES of the C gate) but the magnitude of the LSES and Rabi frequencies is much larger in inhomogeneous strains. In fact, the $g$-factors $|g_\parallel|=0.85$ and $g_\perp=1.17$ are significantly different in the inhomogeneously and biaxially strained devices. This results from the modulations of the confinement potential $\delta E_v^{(\varepsilon)}=-a_v\delta\Omega/\Omega$ and of the HH/LH bandgap $\delta_\mathrm{LH}^{(\varepsilon)}=-2b_v\varepsilon_\mathrm{uni}$ by the inhomogeneous strains (see Fig.~\ref{fig:edges}). In the bulk Ge device, $|\delta_\mathrm{LH}^{(\varepsilon)}|\lesssim 2$\,meV is a significant fraction of the bare HH/LH bandgap $\Delta_\mathrm{LH}\approx 2.5$\,meV opened by vertical confinement (but would be unnoticeable in a strained Ge heterostructure). This reshapes the HH and LH envelopes, with visible fingerprints on the hole densities (see Fig.~\ref{fig:wfns}) and $g$-factors.

The inhomogeneous strains have much more impact on the L than on the C gate (because they do not break the symmetry of the device). The Rabi frequencies and LSES of the L gate are, indeed, enhanced by an order of magnitude. The Rabi oscillations are still dominated by the large modulations of $g_{xz}$ and $g_{zx}$, further promoted by the motion of the dot in the inhomogeneous shear strains $\varepsilon_{xz}$ \cite{Abadillo2023}. The quality factor of the C gate is very low (due to its poor driving efficiency) but the quality factor of the L gate can be even larger than in the fully unstrained device. This results from the existence of sweet lines of the side gates that run near the equatorial plane \cite{Mauro24}. Along these sweet lines, the Larmor frequency of the qubit is little sensitive to lateral electric field fluctuations but the Rabi frequencies are maximal. As discussed in the main text, this is advantageous only if the two-qubit gates can be performed as fast as the single qubit operations. Such devices may, therefore, show optimal performances at small Larmor frequencies, where $f_\mathrm{R}\propto f_\mathrm{L}\propto B$ strongly benefits from the enhancement by inhomogeneous strains but $T_2^*\propto B^{-1}$ can remain long.

The relaxation rates $\Gamma_\mathrm{PH}$ are similar with and without inhomogeneous strains (see Fig.~\ref{fig:gammaph}a). Indeed, the enhancement of the Rabi frequency and LSES results from the motion of the hole in the inhomogeneous shear strains, while the relaxation results from the modulations of the shear strains by the phonons. The latter are the same in homogeneous and inhomogeneous strains (at least in the present bulk phonons approximation).

Similar enhancements of the Rabi frequencies and quality factors by inhomogeneous strains have been previously reported in conventional heterostructures \cite{Abadillo2023,Mauro24}. The peak of the quality factor in Fig.~\ref{fig:Rabi_strains} remains, nevertheless, much broader than in Ge quantum wells grown on Ge$_{0.8}$Si$_{0.2}$ buffers \cite{Mauro24}, despite its apparent thinning. These results highlight that strain management will ultimately become an essential concern in the design of hole spin qubits.

\FloatBarrier

\bibliography{biblio}

\end{document}